\documentclass[aps,pre,twocolumn,preprintnumbers,amssymb,showpacs,superscriptaddress,floatfix]{revtex4-1}
\usepackage{amstext,amssymb}
\usepackage{amsmath}
\usepackage{amsfonts}
\usepackage{xcolor}
\usepackage{graphicx}% Include figure files
\usepackage{dcolumn}% Align table columns on decimal point
\usepackage{bm}% bold math
\usepackage{tikz}
\usetikzlibrary{decorations.pathmorphing,patterns}
\usepackage{color}
\usepackage{bbold}

\newcommand{\rate}{T}
\newcommand{\timeexact}{\mathtt{T}}
\newcommand{\gammaenergy}{\Xi}

\begin{document}
%\begin{CJK*}{GBK}{}
%\preprint{APS}

\title{A Landau-Zener Lindblad equation and work extraction from coherences}
\author{Juzar Thingna}
\email[]{juzar@thingna.uni.lu}
\affiliation{Complex Systems and Statistical Mechanics, Physics and Materials Science Research Unit, University of Luxembourg, L-1511 Luxembourg, Luxembourg}
\author{Massimiliano Esposito}
\email[]{massimiliano.esposito@uni.lu}
\affiliation{Complex Systems and Statistical Mechanics, Physics and Materials Science Research Unit, University of Luxembourg, L-1511 Luxembourg, Luxembourg}
\affiliation{Kavli Institute for Theoretical Physics, University of California, Santa Barbara, CA 93106 Santa Barbara,  U.S.A.}
\author{Felipe Barra}
\email[]{fbarra@dfi.uchile.cl}
\affiliation{Departamento de F\'isica, Facultad de Ciencias F\'isicas y Matem\'aticas, Universidad de Chile, Santiago, Chile}
\affiliation{Kavli Institute for Theoretical Physics, University of California, Santa Barbara, CA 93106 Santa Barbara,  U.S.A.}

\date{\today}

\begin{abstract}
We show that the dynamics of a driven quantum system weakly coupled to a finite reservoir can be approximated by a sequence of Landau-Zener transitions if the level spacing of the reservoir is large enough. This approximation can be formulated as a repeated interaction dynamics and leads to a quantum master equation for the driven system which is of Lindblad form. The approach is validated by comparison with the numerically exact full system dynamics. To emphasize the role of coherence in the master equation, we propose a model-system which shows that in its presence, work can be extracted from a thermal reservoir while if the coherences vanish no work can be extracted. 
\end{abstract}

\maketitle
%\end{CJK*}

%%%%%%%%%%%%%%%%%%%%%%%%%%%%%%%%%%%%%%%%%%%%%%%%%%%%%%%%%%%%%%%%%%%%%%%%%%%%%%%%%%%%%%%%%%%%%%%%%%%%%%%%%%
\section{Introduction}\label{sec:1}

Establishing reliable kinetic descriptions for time-dependently driven open quantum systems is important in many contexts \cite{09KetzmerickKohnPRE, 10TimmPRB, 13HauptWegewijs, 14SilaevVirtanenPRE, 14ThingnaCampisiPRB, 15Zanardi, 16ShiraiNJP, 18DannKosloff}. In recent years this became particularly true in quantum thermodynamics where work has been particulary focused for periodically driven systems \cite{00KohnJSP, 06SegalPRB, 15BulnesEngelEspoNJP, 15GelbwaserNiedenzuKurizki, 15ZhouSciRep, 16LudovicoSanchezArracheaPRB, 16BrandnerSeifertPRE, 17FreitasPazPRE}
but not only \cite{79AlickiJPA, 12HorowitzPRE, 14KosloffEntropy, 15EspoGalperinPRL, 15UzdimLevyKosloffPRX, 16CuetaraEspoSchallerEnt, 18HaughianEspositoSchmidtPRB}. 
We used a distinctive approach in Refs~\cite{Barra16, Thingna17}, where we studied the thermodynamics and dynamics of a quantum dot with a time dependent energy level, coupled to a fermionic reservoir with finite spacing between energy levels. When the coupling is weak compared to that spacing, the dynamics can be described as a sequence of Landau-Zener transitions~\cite{Landau1932, Zener1932, Majorana1932, Stueckelberg1932} occurring whenever the dot energy crosses a reservoir level. The resulting stochastic dynamics for the dot occupation was shown to agree very well with the numerically exact full quantum dynamics.

In these previous studies the initial state was thermal for the reservoir and diagonal in the energy basis of the dot. As a result coherences were absent from the description. 
In the first past of this work, Sec. \ref{sec:2}, we extend these previous works and consider initial states which may contain coherences. We formulate the problem in a different but equivalent physical setup. Our system is now a single spin 1/2 system interacting with a reservoir of $L$ spin 1/2, that eventually will be thermal. We will show that the dynamical description we obtain for the system can be formulated in a repeated interaction framework~\cite{RI, Barra2015, Strasberg2017}. The agreement with numerically exact results will again be shown to be very good.  
In the second part of the paper, Sec. \ref{sec:3}, we use our results to propose a machine that can extract work from coherences, a topic that has attracted attention in recent years~\cite{coh-work, 16JenningsNJP, 17PlenioRMP}. The machine is an autonomous spin 1/2 in permanent contact with a thermal reservoir and subjected to repeated interactions with  driven spin 1/2 systems which are initially prepared in a thermally populated density matrix with non-vanishing coherences. Work extraction in this model is exclusively caused by coherences.
Conclusions are drawn in Sec. \ref{sec:4}.

%%%%%%%%%%%%%%%%%%%%%%%%%%%%%%%%%%%%%%%%%%%%%%%%%%%%%%%%%%%%%%%%%%%%%%%%%%%%%%%%%%%%%%%%%%%%%%%%%%%%%%%%%%
\section{Quantum dynamical map}\label{sec:2}
We consider a spin-$1/2$ particle interacting with a finite number of spins whose Hamiltonian takes the form,
\begin{align}
\label{eq:Hamiltonian}
H(t) &= \varepsilon_t \sigma^z + \sum_{j=1}^{L}\epsilon_j \sigma^z_j + \sum_{j=1}^{L}\nu_j\sigma^+\sigma^-_j+ \mathrm{H.c.}. \\
 &= H_S(t) + H_B + V \nonumber
\end{align}
Above $\sigma^{\alpha}$ ($\alpha = x,y,z$) are Pauli spin-$1/2$ matrices and $\sigma^{\pm}=(\sigma^x \pm i\sigma^y)/2$. Hereon, we refer to the first spin (without subscript) as the system with time dependent $H_S(t)=\varepsilon_t \sigma^z $ and the remaining $L$ spins as the reservoir with $H_B =  \sum_{j=1}^{L}\epsilon_j \sigma^z_j$. When $L\rightarrow \infty$ we obtain a system interacting with a spin reservoir~\cite{Prokofev2000}. 

As the energy level of the system, $\varepsilon_t$, is ramped in time, it will cross the reservoir energy levels $\{\epsilon_j\}$ (see Fig.1 in~\cite{Barra16, Thingna17}) and undergo an avoided crossing in the single spin magnetization (single-particle) basis with an energy gap of order $2\nu_j$. We assume that the level spacing between consecutive reservoir spin energies is greater than that gap, i.e.,
\begin{align}
\epsilon_{j+1}-\epsilon_j > 2|\nu_j|
\label{eq:2}
\end{align}
and that the time between two consecutive levels $\tau_j^c = (\epsilon_{j+1}-\epsilon_j)/\dot{\varepsilon}_j$ is greater than the Landau-Zener validity time $\tau^{lz}_j$~\cite{NoriPR2010}, i.e.,
\begin{align}
\epsilon_{j+1}-\epsilon_j > \sqrt{\dot{\varepsilon}_j}\mathrm{max}\left[1,2|\nu_j|/\sqrt{\dot{\varepsilon}_j}\right].
\label{eq:3}
\end{align}
Here $\dot\varepsilon_j=\dot\varepsilon_t$ at the crossing time $t=\timeexact_j$, i.e., when $\varepsilon_t=\epsilon_j$. Above and throughout this work we will set the Planck constant $\hbar$ and Boltzmann constant $k_B$ to unity. Inequalities Eq.~\eqref{eq:2} and Eq.~\eqref{eq:3} allow us to simplify the problem as follows. First, due to Eq.~\eqref{eq:2}, for any given time $t$, the Hamiltonian can be approximated as,
\begin{align}
\label{eq:Happrox}
H(t) &\approx H_n + H_{\neq n},
\end{align}
with
\begin{align}
H_n &=\varepsilon_t \sigma^z + \epsilon_n \sigma^z_n+\left(\nu_n\sigma^+\sigma^-_n+\mathrm{H.c.}\right), \nonumber \\
H_{\neq n} &= \sum_{j\neq n}\epsilon_j \sigma^z_j
\end{align}
where $n$ is determined by the nearest reservoir energy level $\epsilon_n={\min}_m|\varepsilon_t-\epsilon_m|$. Equivalently $H(t)\approx H_n + H_{\neq n}$ for $t_n<t<t_{n+1}$ with $t_n=(\timeexact_n+\timeexact_{n+1})/2$ and therefore the evolution operator between the times $t_{n-1}$ and $t_{n}$ is thus approximated by 
\begin{align}
U(t_{n},t_{n-1}) &= U_n\otimes U_{\neq n},
\end{align}
with $U_x = \exp\left[-i\int_{t_{n-1}}^{t_{n}}dt H_x\right]$ ($x=n, \neq n$). 
Second, due to Eq.~\eqref{eq:3}, the unitary matrix $U_n$ relating the states of the reduced system-reservoir level described by $H_n$ before and after the avoided crossing is a Landau-Zener transition unitary matrix, see later.
  
The dynamics at this level of approximation takes the form of a repeated interaction problem~\cite{RI, Barra2015, Strasberg2017}, the system interacts with the spins of the reservoir \emph{sequentially}.

Therefore, considering the initial state of the total system as $\varrho_0\otimes \rho^B(0)$, a product state between the initial system density matrix $\varrho_0$ and the initial spin reservoir density matrix $\rho^B(0)=\bigotimes_{j}\rho^B_j$ in which we neglect any correlation between different spins levels but allows for level-coherences (i.e.  $\rho^B_j$ may have non-diagonal elements), 
we have after the first avoided crossing that the system is in the state  
\begin{align}
\varrho_1= {\rm Tr}_0[U_1 \varrho_0\otimes \rho^B_0U_{1}^{\dagger}].
\end{align}
Similarly, at time $t_n$, the system density matrix is
\begin{equation}
\varrho_n={\rm Tr}_{n-1}[U_n \varrho_{n-1}\otimes \rho^B_{n-1}U_{n}^{\dagger}],
\label{eq:dyn}
\end{equation}
where ${\rm Tr}_k$ is the trace over the level $k$ of the bath. This iterative dynamics is valid 
as long as the bare system energy $\varepsilon_t$ is monotonous in time and thus does not cross a reservoir level for a second time. If that were the case, a similar treatment would be possible if the quantum correlations built between the system and reservoir were negligible. 

Thus, the problem of evaluating the quantum dynamical map boils down to obtaining the evolution operator $U_n$. Independent of the number of spins in the reservoir the operator $U_n$ only lives in the Hilbert space of the system and the interacting $n$th reservoir spin. Hence, in order to obtain $U_n= \exp\left[-i\int_{t_n}^{t_{n+1}}dt H_n\right]$ we begin by expressing $H_n$ in the basis that diagonalize $\sigma^z\otimes\sigma^z_n$, which reads,
\begin{align}
\label{eq:Hn}
H_n|\uparrow\uparrow_n\rangle &= (\varepsilon_t + \epsilon_n) |\uparrow\uparrow_n\rangle, \nonumber \\
H_n|\uparrow\downarrow_n\rangle &= (\varepsilon_t - \epsilon_n) |\uparrow\downarrow_n\rangle + \nu_n^* |\downarrow\uparrow_n\rangle, \nonumber \\
H_n|\downarrow\uparrow_n\rangle &= -(\varepsilon_t - \epsilon_n) |\downarrow\uparrow_n\rangle+ \nu_n |\uparrow\downarrow_n\rangle, \nonumber \\
H_n|\downarrow\downarrow_n\rangle &= -(\varepsilon_t + \epsilon_n) |\downarrow\downarrow_n\rangle,
\end{align}
where $\sigma_z|\uparrow\rangle = |\uparrow\rangle$, $\sigma_z|\downarrow\rangle = -|\downarrow\rangle$, and $|\uparrow \downarrow_n\rangle = |\uparrow\rangle \otimes |\downarrow_n\rangle$ with the first ket corresponding to the system spin and the second corresponding to the $n$th reservoir spin. Equation~\eqref{eq:Hn} clearly shows that we get two invariant subspaces that do not influence each other, one corresponding to the single up-spin basis $\{|\uparrow\downarrow_n\rangle, |\downarrow\uparrow_n\rangle\}$, and the other $\{|\uparrow\uparrow_n\rangle, |\downarrow\downarrow_n\rangle\}$. Closer inspection shows that the Hamiltonian $H_n$ in the single up-spin subspace is the same as the standard Landau-Zener problem for a single particle~\cite{Landau1932, Zener1932, Majorana1932, Stueckelberg1932} whereas $H_n$ is diagonal in the other subspace. 

Thus, by virtue of Eq.~\eqref{eq:3} and assuming that the energy of the system spin varies linearly in time, i.e., $\varepsilon_t = \dot{\varepsilon}_n t$ when the system interacts with the $n$th reservoir spin, the evolution operator $U_n$ can be obtained either via the adiabatic impulse method \cite{NoriPR2010} or by contour integration \cite{Wittig2005}. Treating the two invariant subspaces independently, one obtains a Landau-Zener transition unitary matrix,
\begin{align}
\label{eq:Un}
U_n|\uparrow\uparrow_n\rangle &= e^{-i\alpha_n} |\uparrow\uparrow_n\rangle, \nonumber \\
U_n|\uparrow\downarrow_n\rangle &= \sqrt{R_n}|\uparrow\downarrow_n\rangle - \sqrt{1-R_n} e^{i\varphi_n} |\downarrow\uparrow_n\rangle, \nonumber \\
U_n|\downarrow\uparrow_n\rangle &= \sqrt{R_n}|\downarrow\uparrow_n\rangle+ \sqrt{1-R_n} e^{-i\varphi_n} |\uparrow\downarrow_n\rangle, \nonumber \\
U_n|\downarrow\downarrow_n\rangle &= e^{i\alpha_n} |\downarrow\downarrow_n\rangle,
\end{align}
with
\begin{align}
\alpha_n & = \frac{\dot{\varepsilon}_n}{2}\left(t_{n}^2-t_{n-1}^2\right) + \epsilon_n \left(t_{n}-t_{n-1}\right),\\
R_n & = e^{-2\pi \delta_n}, \quad  \delta_n = \frac{|\nu_n|^2}{2\dot{\varepsilon}_n}, \\
\varphi_n &= \frac{\pi}{4} + \delta_n\left(\mathrm{ln}\delta_n -1\right) + \mathrm{arg}\left[\Gamma(1-i\delta_n)\right],
\end{align}
and $\Gamma(x)$ being the Gamma function of $x$. The parameter $\varphi_n$ above is the Stokes phase \cite{Kayanuma97} that is independent of all the details of the evolution and only depends on what happens at the point of interaction where an avoided crossing occurs.
\begin{figure*}[tb!]
\includegraphics[width=\textwidth]{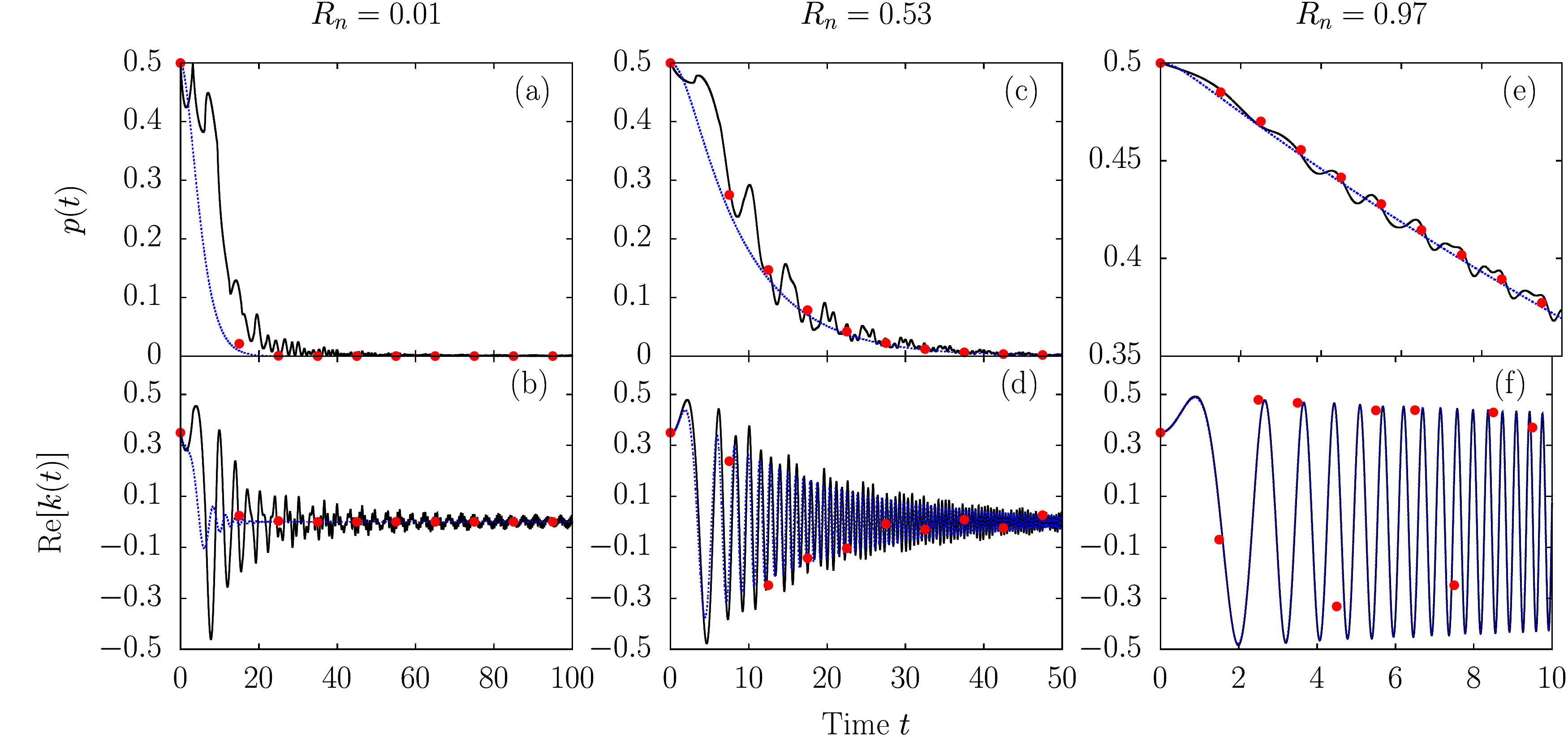}
\caption{\label{fig1} Comparison between exact quantum dynamics (black solid line), Landau-Zener Markov chain [red closed circles, Eqs.~\eqref{eq:popmap} and \eqref{eq:cohmap}], and continuous time Landau-Zener master equation [blue dotted line, Eqs.~\eqref{eq:popctlz} and \eqref{eq:cohctlz}] for different values of Landau-Zener transition probability $1-R_n$. The parameters for panels: (a) and (b) are $\nu_n = 0.4$ and $\dot{\varepsilon}_n = 0.1$, (c) and (d) are $\nu_n = 0.2$ and $\dot{\varepsilon}_n = 0.2$, and (e) and (f) are $\nu_n = 0.1$ and $\dot{\varepsilon}_n = 1.0$. The spin reservoir with $N=10$ is set initially at inverse temperature $\beta = 2$, $\mu=0$, and energies $\epsilon_n = n\epsilon$ with $\epsilon=1$.}
\end{figure*}

Before we proceed, let us summarize the main assumption so far: the weak coupling condition Eq.~\eqref{eq:2} allows us to approximate $H(t)$ as in Eq.~\eqref{eq:Happrox}.  Thus, whenever the system crosses a reservoir level, only the reduced density matrix of the system and that particular reservoir level change. Hence, since the system level moves monotonously  through the sea of reservoir spins levels sequentially it always interacts with the reservoir spin that is in the corresponding state $\rho^B_j$. Moreover, due to Eq.~\eqref{eq:3}, the evolution Eq.~\eqref{eq:dyn} of the spin system is given by the Landau-Zener unitary Eq.~\eqref{eq:Un}. The reduced density matrix of the system at time $t_{n}$ thus reads,
\begin{align}
\varrho_{n} &=  \begin{pmatrix}
p_{n} & k_{n}\\
k^*_{n} & 1-p_{n}
\end{pmatrix},
\end{align}
where the first row and column correspond to the up-spin state $|\uparrow\rangle$ whereas the second row and column belong to the down-spin state $|\downarrow\rangle$. The density matrix at the next time step is determined using Eq.~\eqref{eq:dyn} and Eq.~\eqref{eq:Un} as,
\begin{eqnarray}
\label{eq:popmap-gen}
p_{n+1} &=& R_n p_n + \tilde{a}_n (1-R_n) \nonumber \\
&&+ \sqrt{R_n(1-R_n)}\left[e^{i\varphi_n}\tilde{q}_nk_n^* + \mathrm{H.c.}\right],\\
\label{eq:cohmap-gen}
k_{n+1} &=& \sqrt{R_n} e^{-i\alpha_n}k_n \nonumber \\
&&- \sqrt{1-R_n} e^{-i\left(\alpha_n-\varphi_n\right)}\tilde{q}_n(2p_n-1),
\end{eqnarray}
where 
\begin{align}
\rho^B_{n} &= \begin{pmatrix}
\tilde{a}_{n} & \tilde{q}_{n}\\
\tilde{q}^*_{n} & 1-\tilde{a}_{n}
\end{pmatrix}.
\label{rhoBn}
\end{align}

In the rest of this section we consider the particular case that the reservoir density matrix is initially grand canonical $\rho^B(0)=\exp\left[-\beta(H_B-\mu M_B)\right]/Z_B$ with $M_B =  \sum_{j=1}^{L} \sigma^z_j $ the conserved net magnetization of the reservoir and the parameter $\mu$ the associated spin-chemical potential. Therefore, one has $\rho^B_j=\exp\left[-\beta(\epsilon_j \sigma^z_j -\mu \sigma^z_j )\right]/Z_j,$ i.e., Eq.\eqref{rhoBn} is diagonal with $\tilde{q}_{n}=0$ and $\tilde{a}_{n}=f_n = \{\exp[2\beta(\epsilon_n - \mu)] +1\}^{-1}$ being a modified Fermi-Dirac distribution for spin systems (note the extra factor $2$ with $\beta$) arising due to the initial grand-canonical distribution of the reservoir. In this case, Eq.\eqref{eq:popmap-gen} and Eq.\eqref{eq:cohmap-gen} become
\begin{align}
\label{eq:popmap}
p_{n+1} &= R_n p_n + f_n (1-R_n) \\
\label{eq:cohmap}
k_{n+1} &= \sqrt{R_n} e^{-i\alpha_n}k_n.
\end{align}
The above completely positive trace preserving (CPTP) map forms the first main result of this work. An earlier version of the Landau-Zener master equation was physically motivated and used to address the dynamics of the population only [Eq.(\ref{eq:popmap})] without any reference to the coherence. Our approach has allowed us to provide a unified quantum dynamical map to address populations and coherence on the same footing [Eq.(\ref{eq:popmap}) and Eq.(\ref{eq:cohmap})]. Interesting, for a thermal reservoir, the populations and coherence decouple in the above Markovian dynamical map and the equation for population $p_{n+1}$ matches exactly with the physically motivated discrete time Landau-Zener chain \cite{Barra16}.  It is worth mentioning that the fermionic and spin system, described here, can be mapped onto one another by 
the following mapping: $\sigma_z \rightarrow c^{\dagger}c$, $\sigma^+ \rightarrow c^\dagger$ in the Hamiltonian [Eq.~\eqref{eq:Hamiltonian}] and $\beta \rightarrow \beta/2$, $\dot{\varepsilon}_n \rightarrow \dot{\varepsilon}_n/2$ for the dynamical map in Eqs.~\eqref{eq:popmap} and \eqref{eq:cohmap}. The physical intuition behind this mapping is that in the spin system both the up- and down-spin state energies of the system move in opposite directions giving a net speed of $2\dot{\varepsilon}_n$ at the time of interaction, whereas in the fermionic case only the occupied state energy is time dependent. The reasoning for $\beta$ mapping is similar since in the spin system both up- and down-spin states contribute to the energy and magnetization, whereas in the fermionic case it is only the occupied state contributes.

\subsection*{Continuous time Landau-Zener quantum master equation}
In order to obtain a continuous time Landau-Zener quantum master equation for a system interacting with a thermal reservoir, we consider that during a small interval of time $dt$ the system interacts with $n$ reservoir spins. Thus, by neglecting the variation of minute changes within this small interval the equation for populations takes the form \cite{Thingna17},
\begin{align}
\label{eq:popcontiR}
\frac{dp(t)}{dt} = \dot{\varepsilon}_t \bar{D}_t[1-R_t][f(\varepsilon_t)-p(t)],
\end{align}
with $R_t= \exp\left[-2\pi \delta_t\right]$, $\delta_t = |\nu_t|^2/2\dot{\varepsilon}_t$, and $f(\varepsilon_t) = \{\exp[2\beta(\varepsilon_t - \mu)] +1\}^{-1}$, $\dot{\varepsilon}_t$ being the instantaneous linearized speed of the system at any time $t$. The factor $ \dot{\varepsilon}_t\bar{D}_t$ is an estimation of the number of spins that have interacted with the system in a small interval $dt$ with $\bar{D}_t$ being the reservoir density of states. Next, we transform the coherence map [Eq.~\eqref{eq:cohmap}] into a continuous time version by expressing it in terms of initial condition $k_0=k$ as,
\begin{align}
k_{n+1} &= k\Pi_{j=1}^n\sqrt{R_j}e^{-i\alpha_j} \nonumber \\
\label{eq:cohernecemapinitial}
&= \exp\left[\sum_{j=1}^n\left(\frac{\mathrm{ln}R_j}{2}-i\alpha_j\right)\right],
\end{align}
Using the above equation and noting that in the continuous time limit $t_{j+1}-t_j=dt'$, $t_{j+1}+t_j = 2t'$, and the $\sum_{j=1}^n$ being replaced by $\int_0^t$ we obtain,
\begin{align}
\label{eq:coherencecontiR}
k(t)=k\exp\left[\int_{0}^tdt'\left\{|\dot{\varepsilon}_{t'}|\bar{D}_{t'}\frac{\mathrm{ln}R_{t'}}{2} - 2i\varepsilon_{t'}\right\}\right].
\end{align}
Furthermore, we assume that we work in the diabatic regime [$R_t\approx 1$] and expand $R_t$ in Eqs.~\eqref{eq:popcontiR} and \eqref{eq:coherencecontiR} to obtain,
\begin{align}
\label{eq:popctlz}
d_t p(t) &= \rate^+[1-p(t)]-\rate^-p(t),\\
\label{eq:cohctlz}
d_t k(t)&=-2i\varepsilon_{t}k(t)-\frac{\gammaenergy_{t}}{2}k(t).
\end{align}
with $\rate^+ = \gammaenergy_t f(\varepsilon_t)$, $\rate^- = \gammaenergy_t[1-f(\varepsilon_t)]$, and $\gammaenergy_t = \pi\bar{D}_{t}|\nu_{t}|^2$. The above equations are of Lindblad form and match the Markovian Redfield equation \cite{Redfield1957} after we neglect the Lamb shifts. The coherence oscillate in time, due to the coherent evolution, but at long times they decay with a rate $\gammaenergy_{t}$.

The comparison between the exact quantum dynamics obtained by solving the Schr\"{o}dinger equation and the approximate Landau-Zener theory is found in Fig.~\ref{fig1}. The thermal reservoir here consists of only $10$ spins with equally spaced energy levels $\epsilon_n = n\epsilon$ such that the reservoir has a uniform density of states  $\bar{D}_t = \epsilon^{-1}$. The Landau-Zener Markov chain depicted as red closed circles [Eqs.~\eqref{eq:popmap} and \eqref{eq:cohmap}] matches well with the exact dynamics (black solid lines) irrespective of the Landau-Zener transition probability $1-R_n$ for both populations [Fig.~\ref{fig1}(a), (c), and (e)] and coherence [Fig.~\ref{fig1}(b), (d), and (f)]. In all cases the Markov chain for the coherence shows a slight deviation from the exact result because of the fast oscillations of the coherence in between the two avoided crossings. Surprisingly, these oscillations, even in between two avoided crossings, are fully captured by the continuous time approach [Eqs.~\eqref{eq:popctlz} and \eqref{eq:cohctlz}] in the regime of fast driving when $R_n$ is close to one [Fig.~\ref{fig1}(e) and (f)]. In the slow driving regime [Fig.~\ref{fig1}(a) and (b)] the continuous time approach performs the worst since the key assumption to derive the continuous time version, $R_n \approx 1$, is severely violated. Overall, within its regime of applicability the Landau-Zener approach is able to perfectly mimic the exact quantum dynamics for both populations and coherence. Here we are limited by the computational resources to go beyond $10$ spins in the reservoir, but as seen for the populations~[\onlinecite{Thingna17}] we expect a much better agreement between the Landau-Zener approach and the exact quantum dynamics as the number of spins in the reservoir increases.

\section{Work extraction using coherences}\label{sec:3}
\begin{figure}[t!]
\begin{tikzpicture}[scale=1]
\draw[line width=1pt,->, draw=black] (-0.5,0) -- (6,0);
\draw[line width=1pt,->, draw=black] (-0.5,0) -- (-0.5,5);
\draw[->,line width=1pt] (4.61111,2.5) -- (4.81111,1.825)node[below, align=center] {LZ \\ transition};
\filldraw[fill=white, draw=red, dashed, thick] (4.61111,2.5) circle (0.2cm);
\filldraw[fill=blue!40,draw=blue,dashed,thick] (0.71111,2.4) rectangle (2.51111,2.6);
\node[above,align=center,text=blue] at (2.21111,3.175) {System \\ Redfield \\ dynamics};
\node[above, align=center] at (0.6,0.15) {Driven \\ atoms};
\node[left, align=center, rotate=90] at (-1.3,3.5) {Autonomous \\ system};
%\draw[->,line width=1pt] (1.61111,2.6) -- (2.21111,3.175)node[above, align=center] {System \\ Redfield \\ dynamics};
%\draw[->,line width=1pt] (-0.5,0.1) -- (0.3,0.3)node[above, align=center, anchor=south, xshift =0.2cm] {Driven \\ atoms};
%\draw[->,line width=1pt] (1.5,0.1) -- (0.75,0.3);
%\draw[->,line width=1pt] (-0.4,2.5) -- (0.2,3.8)node[above, align=center, anchor=south, xshift =0.2cm] {Autonomous \\ system};
%
\draw[line width=1pt, dotted]  (1.5,4.5) -- (1.5,0);
\draw[line width=1pt, dotted]  (3.5,4.5) -- (3.5,0);
\draw[line width=1pt, dotted]  (5.5,4.5) -- (5.5,0);
\draw[line width=1pt, dotted] (0.31111,2.5) .. controls (0.61111,2.5) .. (0.91111,3.175);
\draw[line width=1pt, dotted] (0.31111,1.825) .. controls (0.61111,2.5) .. (0.91111,2.5);
\draw[line width=1pt] (-0.5,2.5)node[left] {$\varepsilon_{0}$} -- (0.51111,2.5)node[below,midway,rotate=66, xshift=-0.8cm]{$\dot{\varepsilon}_n t$} ;
\draw[line width=1pt, dotted] (-0.5,0)node[left] {$0$} -- (0.31111,1.825);
\draw[line width=1pt, dotted] (0.91111,3.175) -- (1.5,4.5);
\draw[line width=1pt] (0.71111,2.5) -- (1.5,2.5);
\draw[line width=1pt, dotted] (2.31111,2.5) .. controls (2.61111,2.5) .. (2.91111,3.175);
\draw[line width=1pt, dotted] (2.31111,1.825) .. controls (2.61111,2.5) .. (2.91111,2.5);
\draw[line width=1pt] (1.5,2.5) -- (2.51111,2.5);
\draw[line width=1pt, dotted] (1.5,0)node[below] {$T$} -- (2.31111,1.825);
\draw[line width=1pt, dotted] (2.91111,3.175) -- (3.5,4.5);
\draw[line width=1pt] (2.71111,2.5) -- (3.5,2.5);
\draw[line width=1pt, dotted] (4.31111,2.5) .. controls (4.61111,2.5) .. (4.91111,3.175);
\draw[line width=1pt, dotted] (4.31111,1.825) .. controls (4.61111,2.5) .. (4.91111,2.5);
\draw[line width=1pt] (3.5,2.5) -- (4.51111,2.5);
\draw[line width=1pt, dotted] (3.5,0)node[below] {$2T$} -- (4.31111,1.825);
\draw[line width=1pt, dotted] (4.91111,3.175) -- (5.5,4.5);
\draw[line width=1pt] (4.71111,2.5) -- (6,2.5);
\draw[line width=0pt] (5.5,0) -- (5.5,0) node[below] {$3T$};
\filldraw[fill=black!40, draw=black, thick] (-0.4,2.5) circle (0.12cm);
\filldraw[fill=black!40, draw=black, thick] (-0.5,0.1) circle (0.12cm);
\filldraw[fill=black!40, draw=black, thick] (1.5,0.1) circle (0.12cm);
\filldraw[fill=black!40, draw=black, thick] (3.5,0.1) circle (0.12cm);
\end{tikzpicture}
\caption{\label{fig2} An illustration of the work machine that utilizes coherences from the atoms to extract work. The energy of the atoms is linearly driven across an autonomous system that evolves connected to a thermal reservoir. The resonant coupling between the atoms and the system allows for a Landau-Zener transition that leads to an exchange in populations and coherence given by the quantum dynamical map.}
\end{figure}
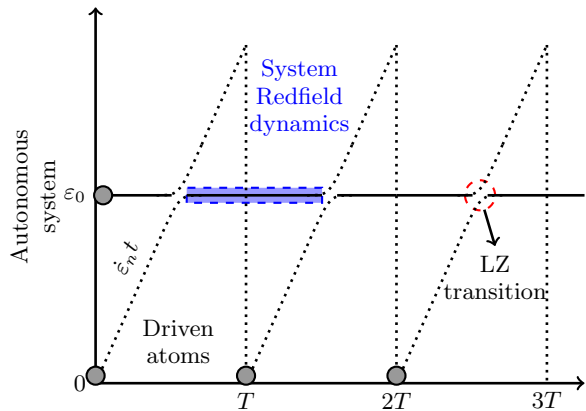
In this section we study the role of coherences in steady state work extraction using a simple model of an autonomous system $H_S$ connected to a thermal reservoir $H_R$ and interacting with two-level atoms $H_A(t)$. The linearly driven atoms resonantly interact with the system as illustrated in Fig.~\ref{fig2}. The Hamiltonian reads
\begin{eqnarray}
H(t) &= H_S + H_A(t) + H_R + H_{SR} + H_{SA}
\end{eqnarray}
with
\begin{align}
H_S &= \varepsilon_0 \sigma^z ,  \quad H_A (t)= \sum_j \varepsilon_j^t \sigma_j^z , \nonumber \\
H_R &= \sum_k \frac{p_k^2}{2 m_k}+ \frac{1}{2}m_k\omega_k^2 x_k^2 , \nonumber \\
H_{SR} & = (\sigma^x+\sigma^z)\sum_k c_k x_k , \nonumber \\
H_{SA} &= \sum_j \nu_j \sigma^+ \sigma^-_j + \mathrm{H.c.}.
\end{align}
The reservoir is composed of thermal harmonic oscillators with mass $m_k$ and frequencies $\omega_k$ that couple to the system causing energy changes ($\sigma^z$ coupling) and spin flips ($\sigma^x$ coupling) in the system simultaneously. Without the thermal reservoir the Hamiltonian would be similar to Eq.~\eqref{eq:Hamiltonian} with the system now being autonomous and the reservoir driven. 

When coupled only to the reservoir (i.e. when $H_{SA}=0$), the system density matrix $\varrho$ follows the Markovian Redfield quantum master equation \cite{Redfield1957, Breuer2002} that, given the form of the term $H_{SR}$, couples population and coherence and is given by,
\begin{align}
\label{eq:Redfield}
d_t \varrho_{nm} = & -i\Delta_{nm}\varrho_{nm} +\sum_{i,j}\mathcal{R}^{ij}_{nm}\varrho_{ij}, \\ 
\mathcal{R}_{nm}^{ij} = & S_{ni}S_{jm}\left(W_{ni}+W_{mj}\right)-\delta_{j,m} \sum_{l} S_{nl}S_{li}W_{li} \nonumber \\
& -\delta_{n,i} \sum_{l} S_{jl}S_{lm}W_{lj}, \nonumber
\end{align}
with the non-zero elements of $\Delta_{nm}$ being $\Delta_{12} = -\Delta_{21} = 2\varepsilon_0$. The operator $S = \sigma^x +\sigma^z$ has elements $S_{11}=S_{12}=S_{21}=-S_{22}=1$, and the Markovian rates read $W_{12} = J(2\varepsilon_0)n(2\varepsilon_0)$, $W_{21} = J(2\varepsilon_0)\left[n(2\varepsilon_0)+1\right]$ and $W_{11} = W_{22} = \lim_{x\rightarrow 0} J(x)n(x)$ (the Lamb-shift are ignored). Also, $J(\omega) = \eta \omega/\left[1+\left(\omega/\omega_c\right)^2\right]$ is an ohmic spectral density with a Lorentz-Drude cutoff $\omega_c$ and $n(\omega) = \left[\exp(\beta\omega)-1\right]^{-1}$ is the Bose-Einstein distribution. The energy change of the system in a given interval of time $(t,t+\tau)$, due to the contact with the reservoir is associated to a heat flow, i.e., $Q={\rm Tr}_S [H_S[\varrho(t+\tau)-\varrho(t)]]$. 

Consider now the effect of $H_{SA}$ but similar to the assumptions imposed in the previous section we assume $\varepsilon_{j+1}^t-\varepsilon_j^t>2\nu_j$. In this regime, the effect of the system-atom interaction is important only for a short interval of time around the crossing times $\timeexact_n$ given by $\varepsilon_n^{\timeexact_n}=\varepsilon_0$ (see Fig.~\ref{fig2}). Between these crossings, i.e., for times $\timeexact_n<t<\timeexact_{n+1}$ we neglect the effect of $H_{SA}$ and thus the system obeys Redfield dynamics. At the crossing times $\timeexact_n$ the density matrix of the system and the atom change instantaneously with a map provided by the Landau-Zener theory described in Sec.~\ref{sec:2}. In particular, the map for the system takes the form of Eq.~\eqref{eq:dyn} with the $U_n$ as in Eq.~\eqref{eq:Un} with the replacements $\epsilon_n \rightarrow \varepsilon_0$ and $\varphi_n \rightarrow -\varphi_n$.

\begin{figure*}[tb!]
\includegraphics[width=2\columnwidth]{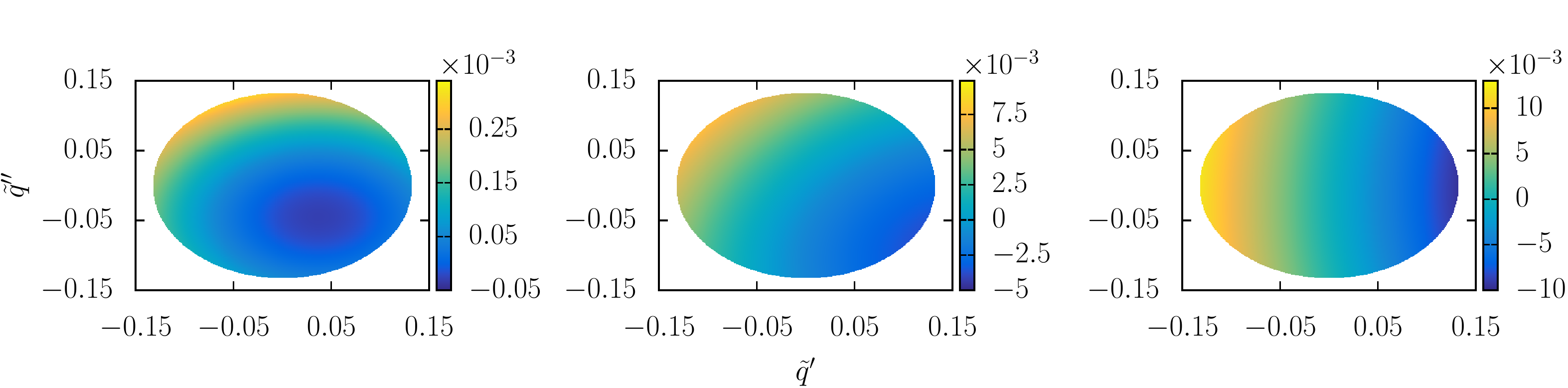}
\caption{\label{fig3} Power (mechanical work per unit cycle), $P_{mech} = W_{mech}/T$, as a function of the initial coherence $\tilde{q}_n=\tilde{q}^{\prime}+ i \tilde{q}^{\prime\prime}$ of the atoms. The panels represent different Landau-Zener interaction strengths $\nu = 0.1$ (left panel), $\nu=0.45$ (middle panel), and $\nu=1$ (right panel) that affect the Landau-Zener transition probability $R_n = R \approx 1, 0.5, 0$ from left to right. The common parameters used are: $\varepsilon_0 = 1$, $\gamma = 0.1$, $\beta = \beta_{LZ} = 2$, $\mu=0$, $\omega_c = 10$, $T = 2$, and $\dot{\varepsilon} = 1$.}
\end{figure*}

The system density matrix just before  $\timeexact_n$ contain both populations and coherence due to the Redfield evolution and is taken of the form
\begin{align}
\varrho_{n-1} &=  \begin{pmatrix}
p_{n-1} & k_{n-1}\\
k^*_{n-1} & 1-p_{n-1}
\end{pmatrix}.
\end{align}
The initial density matrix of the $n$th atom is designed to contain populations that are thermal with respect to the harmonic oscillator reservoir and coherence such that the matrix reads
\begin{align}
\rho^a_{n} &=  \begin{pmatrix}
\tilde{a}_{n} & \tilde{q}_{n}\\
\tilde{q}^*_{n} & 1-\tilde{a}_{n}
\end{pmatrix}
\end{align}
with $\tilde{a}_n =  \{\exp[2\beta(\varepsilon_0 - \mu)] +1\}^{-1}$ [see below Eq.~\eqref{eq:cohmap}]. Here $\tilde{a}_n$ is chosen to be independent of $n$ with a thermal distribution w.r.t. the energy of the autonomous system $\varepsilon_0$. This choice is not due to a restriction of the theory. It allows us to focus on the effect of coherence, since in their absence the environment of the system is thermal and work per cycle would be zero (as explained later). 
Now, similarly to the process that leads to Eqs.\eqref{eq:popmap-gen} and \eqref{eq:cohmap-gen}, 
applying the evolution operator on the decoupled density matrix between system and atom, i.e., $U_{n+1} \varrho_n \otimes \rho^a_n U_{n+1}^{\dagger}$ and keeping track of both system and atom after interaction [tracing over the atom and system respectively] we obtain,
\begin{eqnarray}
\label{eq:popsmap}
p_{n+1} &=& R_n p_n + \tilde{a}_n (1-R_n) \nonumber \\
&&+ \sqrt{R_n(1-R_n)}\left[e^{i\varphi_n}\tilde{q}_nk_n^* + \mathrm{H.c.}\right],\\
\label{eq:cohsmap}
k_{n+1} &=& \sqrt{R_n} e^{-i\alpha_n}k_n \nonumber \\
&&- \sqrt{1-R_n} e^{-i\left(\alpha_n-\varphi_n\right)}\tilde{q}_n(2p_n-1),\\
\label{eq:popamap}
a_{n+1} &=& R_n \tilde{a}_n + p_n (1-R_n) \nonumber \\
&&- \sqrt{R_n(1-R_n)}\left[e^{i\varphi_n}\tilde{q}_nk_n^* + \mathrm{H.c.}\right],\\
\label{eq:cohamap}
q_{n+1} &=& \sqrt{R_n} e^{-i\alpha_n}\tilde{q}_n \nonumber \\
&&+ \sqrt{1-R_n} e^{-i\left(\alpha_n+\varphi_n\right)}k_n(2\tilde{a}_n-1)
\end{eqnarray}
with $a_{n+1}$ and $q_{n+1}$ denoting the populations and coherence of the atom after the interaction. 
At the start of each cycle we have a new atom interacting with the system that periodically pumps coherence due to the Landau-Zener transition. These coherence are then mixed with the populations due to the Redfield dynamics. Hence, one expects the past coherence to affect the steady-state populations of both the system and atoms after the avoided crossing. 

The natural question that arises is what is the effect of these initial coherences on steady-state work and if they could be used as a resource to produce work. The mechanical work in one cycle of period $T$ (see Fig.~\ref{fig2}) can be decomposed into three parts: 1) From time $nT$ to the avoided crossing, the mechanical work is the energy change of the atom with the initial state remaining unchanged and is given by,
\begin{align}
W_{mech}^1 & = \varepsilon_0 (2\tilde{a}_n -1).
\end{align}
2) From after the avoided crossing up-to time $(n+1)T$ when the atom energy is at its maximum the mechanical work is given by,
\begin{align}
W_{mech}^2 & = \varepsilon_0 (2a_{n+1}-1),
\end{align}
wherein $a_{n+1}$ [Eq.~\eqref{eq:popamap}] implicitly contains information about all previous coherence due to Redfield evolution that mixes the populations and coherence. Above we have considered the maximum energy of the atom to be $2\varepsilon_0$. 
3) At time $(n+1)T$ when the atom energy drops from its maximum to zero closing the cycle, the state of the atom doesn't change even though it crosses the system energy due to the infinite driving speed. The contribution to the work during this part of the process is
\begin{align}
W_{mech}^3 & = -2\varepsilon_0 (2a_{n+1}-1).
\end{align}
Thus, the mechanical work done over one cycle is
\begin{eqnarray}
\label{eq:Wmech}
W_{mech} &=& W_{mech}^1+W_{mech}^2+W_{mech}^3, \nonumber \\
& =& 2\varepsilon_0(\tilde{a}_n - a_{n+1}),
\end{eqnarray}
which is minus the energy change in the atom due to the crossing, i.e., $W_{mech}={\rm Tr}_A(\varepsilon_n^{\timeexact_n}\sigma^z_n[\rho_{n}^a-\rho_{n+1}^a])$. Work can either be done ($W_{mech} > 0$) or extracted ($W_{mech} < 0$) and in the following for definiteness we will discuss about the extracted work, although the same arguments hold for the work done. The work extracted by manipulating the energy of the atom comes from the reservoir. This is because, since $[U_{LZ},H_S+\varepsilon_n^{\timeexact_n}\sigma^z_n]=0$, one finds 
$W_{mech}=\Delta E_S={\rm Tr}_S(H_S[\varrho_{n+1}-\varrho_n])$. Moreover, since the system reaches a periodic steady state, the energy change of the system over the full period vanishes, i.e., $\Delta E_S+Q=0$, where $Q$ is the heat flow from the (Redfield) reservoir to the system in the period $T$. Thus, the extracted work is the heat flowing from the Redfield reservoir to the system and the coherence in the atom facilitate to pump this heat. In the absence of coherence $k_n=\tilde{q}_n=0$, the steady-state populations for the Redfield dynamics, due to the particular choice of $\tilde{a}_n$ being thermal, would be $p_n=\tilde{a}_n=a_{n+1}$ leading to $W_{mech}=0$ and hence no heat flow.

In order to study the effect of coherences we plot the steady-state mechanical power $P_{mech} = W_{mech}/T$ per unit cycle in Fig.~\ref{fig3} for different values of Landau-Zener interaction strength $\nu_n = \nu$. We consider each atom to be identically prepared such that the initial coherence of the atoms can be split into real and imaginary parts, i.e., $\tilde{q}_n = \tilde{q} = \tilde{q}' + i \tilde{q}''$. The uncolored (white) parts of Fig.~\ref{fig3} is the inaccessible region, since for these values of $q'$ and $q''$ the initial density matrix of the atom is no longer positive.  
The simulations indicate that the power profile is not symmetric with respect to the real and imaginary parts of the initial coherence and the skewness results in a higher work extraction for positive real parts. The maximum work is extracted when the Landau-Zener interaction is the strongest due to a high probability of exchange $1-R_n$ between the system and the atom.

\begin{figure}[tb!]
\includegraphics[width=\columnwidth]{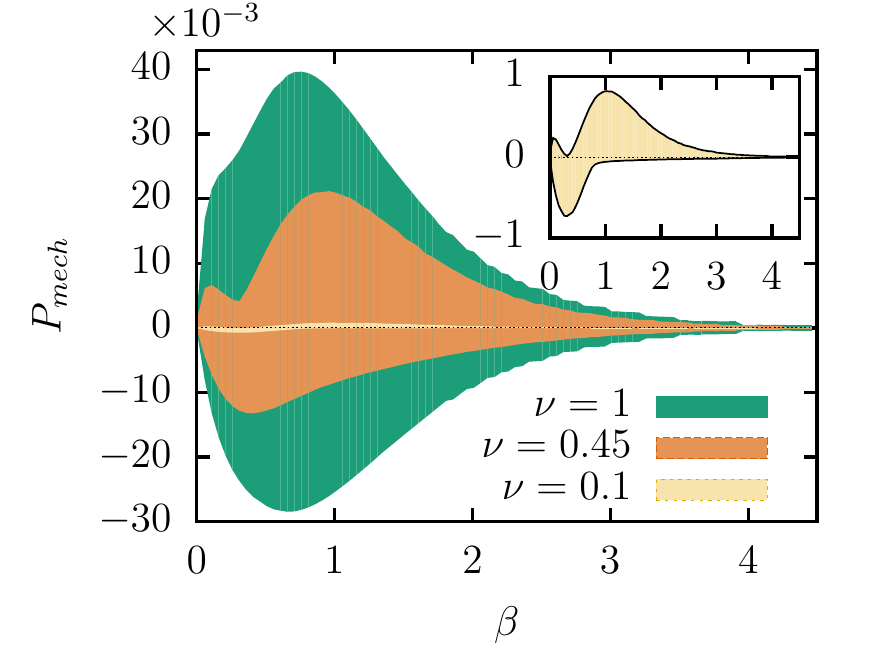}
\caption{\label{fig4} Range of mechanical power, plotted as the shaded region, between maximum and minimum power as a function of inverse temperature $\beta$ for the same parameters as in Fig.~\ref{fig3}.}
\end{figure}
As the temperature $\beta$ is varied, the populations of the atoms vary and hence the accessible range of coherence $\tilde{q}^{\prime}$ and $\tilde{q}^{\prime\prime}$, that ensures that the initial density matrix of the atom is positive, changes. Thus, the global maximum mechanical work done ($P_{mech} > 0$) and extracted ($P_{mech} < 0$) per cycle taken with respect to the real and imaginary part of coherence would vary as a function of $\beta$. In Fig.~\ref{fig4} we plot the entire range of mechanical power $P_{mech}$ accessible at each value of the inverse temperature $\beta$. The maximum power is the maximum work performed and the minimum power (If $< 0$) is the maximum extracted work with the color shade in between representing all the other attainable values of mechanical power for a given $\beta$. This temperature dependence of the mechanical power brings out another interesting asymmetry for the maximum extracted work and work done that occurs at weak to moderate Landau-Zener interactions. In this regime, the maximum work extracted from the atoms occurs in the high temperature (low $\beta$) regime, whereas the maximum work done is in the low temperature regime. This could be a possible control strategy to tune the machine to either extract or do work depending only on the temperature of operation. At strong Landau-Zener interactions this asymmetry disappears and the maximum extracted and done work both occur close to $\beta =\varepsilon_0$.

\section{Conclusions and Discussion}\label{sec:4}

In this paper we extended the Landau-Zener master equation studied in~\cite{Barra16, Thingna17} by incorporating the coherence dynamics and showing that the resulting quantum master equation is of Lindblad form. The main idea is to approximate the system-reservoir interaction as a repeated interaction problem where every interaction is described as a Landau-Zener crossing. 
We showed that the theory agrees very well with the numerically exact full quantum simulations.  
To illustrate the theory and the role of coherences, we proposed a toy model which shows that coherences in the working fluid allow work extraction from a thermal bath.
While presented on a specific model, the method used to derive our Landau-Zener quantum master equation should be generalizable to any noninteracting open quantum system. The extension to interacting models is an interesting future research avenue.  

\begin{acknowledgments}
JT and ME are supported by the European Research Council (Project No. 681456). FB gratefully acknowledges the financial support of FONDECYT grant 1151390 and by the Millennium Nucleus ``Physics of active matter'' of the Millennium Scientific Initiative of the Ministry of Economy, Development and Tourism (Chile). This work was also supported by the National Science Foundation under Grant No. NSF PHY-1748958. 
\end{acknowledgments}

%------------------------------------------  Reference -----------------------------------------------------%
%\bibliography{References}% Produces the bibliography via BibTeX.
%
\end{document}